\documentclass[journal]{IEEEtran}
\usepackage{amssymb}
\usepackage{amsmath}
\usepackage{multirow}
\usepackage{txfonts}
\usepackage{multicol}
\usepackage{booktabs}
\usepackage{mathdots}
\usepackage{graphicx} 
\usepackage{fancyhdr}
\usepackage{hyperref}
\usepackage{float}
\usepackage{microtype}
\usepackage{caption}
\usepackage[numbers]{natbib}

\newcommand{\etal}{{\it et~al.}}

\usepackage{amsmath, scalerel}

\usepackage[linesnumbered,ruled]{algorithm2e}
\usepackage{graphicx}
\graphicspath{ {images/} }
\usepackage{pgfplots, lipsum}
\usepackage{textcomp}
\pgfplotsset{compat = newest}

\pgfplotsset{width=10cm,compat=1.9}
\usepgfplotslibrary{external}
\usepackage{tikz}
\usetikzlibrary{arrows}
\usepackage{caption}
\usepackage{siunitx}
\usepackage[font={small}]{caption}
\usepackage{amsmath}
\usepackage[export]{adjustbox}
\usepackage{lipsum}
\usepackage{subcaption}

\usepackage{xcolor}
\usepackage{pgfplots}
\usepackage{tikz}
\usepackage{pgfkeys}
\usepackage{url}
\usepackage{pgfplotstable}
\usepackage{subcaption}
\usepackage{filecontents}
\usepackage{multirow}
\usepackage{pgf}

\ifCLASSINFOpdf
\else
\fi
\begin{document}
%
\title{Cost Minimization of Cloud Services  for On-Demand Video Streaming}
%
%
%
\author{\IEEEauthorblockN{Mahmoud Darwich$^{*}$, Yasser Ismail$^{2}$}, Talal Darwich$^{3}$, Magdy Bayoumi$^{4}$\\ 
\IEEEauthorblockA{ $^{*}$Department of Mathematical and Digital Sciences, 
Bloomsburg University of Pennsylvania, PA 17815\\
$^{2}$Electrical Engineering Department, 
Southern University and A\&M College, Baton Rouge LA 70807\\
$^{3}$Microchip Technology Inc., San Jose, CA 95134\\
$^{4}$Department of Electrical and Computer Engineering,
University of Louisiana at Lafayette, LA 70504 \\
Email: $^{*}$mdarwich@bloomu.edu, yasser\_ismail@subr.edu, talal.darwich@microchip.com, magdy.bayoumi@louisiana.edu\\
$^{*}$Phone: (570)389-4500, $^{*}$ORCID: 0000-0001-6337-5596
}
}
\maketitle

\begin{abstract}
Cloud Technology is adopted to process video streams because of the great features provided to video stream providers such as the high flexibility of using virtual machines and storage servers at low rates. Video stream providers prepare several formats of the same video to satisfy all users' devices specification. Video streams in the cloud are either transcoded or stored. However, storing all formats of videos is still costly. In this research, we develop an approach that optimizes cloud storage. Particularly, we propose a method that decides which video in which cloud storage should be stored to minimize the overall cost of cloud services. The results of the proposed approach are promising, it shows effectiveness when the number of frequently accessed video grow in a repository, and when the views of videos increases. The proposed method decreases the cost of using cloud services by up to 22\%.
\end{abstract}


\section{Introduction}\label{introduction}
Video streaming became widely accessible by all display devices. Almost every house around the globe possesses at least a smartphone or desktop computer, or laptop, which eases accessing social media to upload and post video streams\cite{lisurvey, report}. Users usually post several videos on social media. Moreover, a video stream needs to be converted to several formats to meet the specifications of different display devices (e.g., network bandwidth, frame rate, and resolution) \cite{ahmad}. Thinking about the huge number of users who post videos on the internet is challenging and needs to be addressed.

Generally, Video conversion requires powerful machines to perform the transcoding operations promptly to avoid any delay on users' end. Therefore, Video stream providers adopted the cloud platform to transcode their videos. They prepare several formats and stored them to make them ready upon request by the users. However, storing all formats incurs a high cost paid by the video stream providers \cite{livlsc}.

 The pattern of accessing video streams is a long-tail distribution. This means the beginning of the curve is allocated for frequently accessed video streams and the remaining of the curve is for rarely accessed videos. The frequently accessed video streams are a small portion while the rarely accessed video streams are huge. The access pattern implies that the video stream providers' repositories contain a small percentage of frequently accessed videos and a large number of rarely accessed videos \cite{sharma}.

Transcoding operation is intensive and costly, it is more expensive than storage cost because the virtual machines are charged on an hourly basis, and storing servers are charged monthly \cite{amazon}. Transcoding all formats and videos could increase the cost significantly when using cloud services. Therefore, it would be less expensive if the video stream providers find a method to store the videos \cite{li1}.

Accordingly, storing the frequently accessed video streams is an effective solution to avoid the costly transcoding operation each time a video is accessed. In this paper, we discuss an approach to how to store the frequently accessed video in cloud storage.
Particularly, we develop a method that clusters the frequently accessed video streams and then stores them in their suitable cloud storage.

Our contributions in this research are summarized as follows: \textbf{(A)} we propose an approach that reduces the cost of could storage by applying clustering on frequently accessed videos. \textbf{(B)} we analyze the effectiveness of the approach when changing the percentage of frequently accessed videos in repositories and when changing the videos views.

This research is an extended work of the previous research \cite{darwich20}. Particularly, the cost of transcoding a video stream is modeled in section \ref{tc}. Furthermore, we develop a model (ratio) that decides either storing or deleting and transcoding of a video stream in the repository as discussed in section \ref{decision}. Thus, the cost of transcoding and the ratio are considered in the execution of the proposed algorithm (Steps 1-10) as detailed in section \ref{algorithm}. Then we implement experiments and analyze the effect of the proposed approach for different scenarios in terms of views and percentage of frequently accessed video variation. Accordingly, we analyze the performance of the proposed method when the views of video vary as illustrated in Fig.\ref{views}. Additionally, we analyze the effectiveness of the proposed method when we include the CDN cost in the experiments as shown in Fig.\ref{cdn}. 

The structure of this paper is organized as follows: Section \ref{introduction} introduces the research topic, Section \ref{related-work} discusses the related work, Section \ref{background} provides a background, Section \ref{proposed} presents the proposed method, Section \ref{experiment} shows the experiment setup, Section \ref{results} reveals the results discussion, and Section \ref{conclusion} concludes the paper.

\section{Related Work}\label{related-work}

Several related works were done to reduce the incurred cost of cloud paid by video companies. Jin \etal \cite{jin2019} presented an erasure frame coding for video streams in the cloud to categorize the videos based on their importance. Their experimental result shows the effectiveness of their design and the reduction of the storage cost. Balamurugan \etal \cite{bala2021} proposed an algorithm that aims to eliminate duplicate video without affecting the unique video data for video surveillance systems. Their proposed algorithm achieves a high efficiency of saving more storage in the system and thus it reduces the storage cost  Haynes \etal \cite{haynes2021} developed a system for video storage that improves the efficiency of storing streaming by up to 45\% and reduces the storage cost by up 54 \%. Their scheme manages video data and eliminates video redundancies from the storage system. Yan \etal \cite{yan2018} proposed a method that removes the duplicate video versions from the cloud storage while maintaining the privacy of IoT metadata. Their proposed paradigm operates at the file-level and deduplication levels. Hu \etal \cite{hu2017} addressed the problem of sharing many videos on social media which poses a high pressure on the cloud storage infrastructure. The authors tackled this problem by proposing a model that motivates the users to upload videos to local storage. Their approach reduces the traffic and service latency.  Kim \etal \cite{kimmultimedia2019} developed an approach by implementing an intra-cloud and parallel framework to transcode videos. Their design could improve precisely the tasks assignments on virtual machines and reduced the transcoding time of videos. Iturriaga \etal \cite{santiago2019} presented a scheme that distributes the content of video streams in cloud-based Content Delivery Network CDN. their proposed approach minimized the utilization of cloud resources. Thus, the incurred cost was reduced while maintaining the quality of service QoS of the users. Darwich \etal \cite{darwichhotness} developed algorithms aiming to decrease the cost of the cloud resources paid by videos companies. They came up with an approach that calculates the access frequency to a video and thus their proposed algorithms categorize the videos that should be stored or deleted and then transcoded upon request. Gao \etal \cite{gao} developed a method that applies the transcoding operation on parts of videos. Their scheme stores the first part of a video which is more frequently accessed while deleting the rest of it and then transcoding it upon request. Their results came promising where they could efficiently use the cloud resources and lower the cost. They proved that their model could reduce the overall cost by up to 30\% compared to fully storing the video. Zhao \etal \cite{zhao} presented an approach that trades-off between the transcoding operations and storing videos to minimize the cost of using the cloud. Particularly, they built a weight graph that describes the transcoding relations among videos. they used the views and the weight graph to decide which videos should be stored or deleted. Their results showed a cost reduction compared to previous methods. Jokhio \etal \cite{jokhio} proposed a scheme that decides whether the video is to be stored or deleted and transcoded on-demand. In their design, they input the video information popularity to their algorithm. Then their method decides how long the video should be kept in the repository. Their results revealed efficiency in terms of cost.

\section{Background}\label{background}
\subsection{ Content of Video Stream }
\begin{figure}[H]
  \includegraphics[width=8cm, height=6cm]{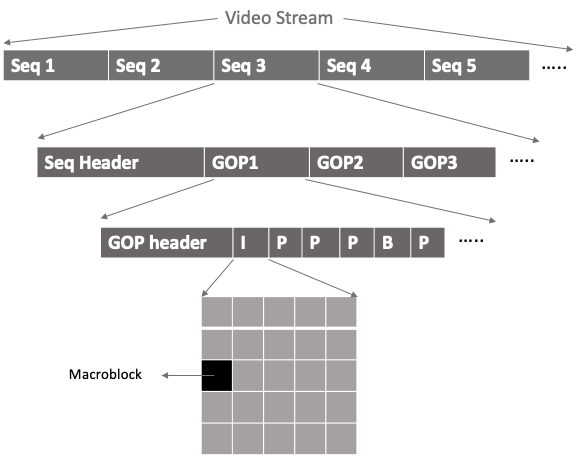}
  \caption{Video Stream Structure}
  \label {video-stream-structure}
\end{figure}

The content of a video stream is formed by several sequences as shown in Fig.~\ref{video-stream-structure}. Each sequence is composed by \emph{Group Of Pictures (GOPs)}. A sequence starts with a block called header which contains data about the sequence. Further, a GOP consists of several frames, it starts with a header, and then a sequence of different types of frames are followed. P (Predicted), B (Bi-directional), and I (intra) frames are the major types found in a GOP. Further, a frame contains small blocks called macroblock (MB) \cite{jokhio1}.

Video streams are created with different formats based on bit rate, spatial resolution, and frame rate. Generally, before streaming a video, it is modified to meet the user's device standards (e.g., frame rate, network bandwidth, and device resolution). These modification operation are achieved on GOPs \cite{intro_6}, \cite{intro_7} and is called video transcoding. we brief the transcoding types as follows:

\subsubsection{Bit Rate Adjustment}
High-quality videos are streamed with a high bit rate and require large bandwidth. Taking into consideration the variation of the bandwidth among users' devices, a video is modified based on the requirement of the network bandwidth to avoid any delay on the user's side~\cite{bg_3}.
\subsubsection{Spatial Resolution Reduction}
Considering the variety of users' devices, a video stream does not fit all display screens specification. Therefore, the spatial resolution should be adjusted and downscaled to match low-resolution screens. In such cases, the quality of videos is impacted and reduced~\cite{bg_7}.
\subsubsection{Video Compression Standard Conversion}
Video compression techniques are highly important for streaming and storing because some videos have large sizes and need to be decreased. The common used compression techniques are MPEG2, H.264, and HEVC~\cite{haskell1996digital},~\cite{wiegand2003overview},~\cite{intro_3}.

\subsection{Cloud Services Video Stream Transcoding}
Cloud resources are well known for the great services offered to users because they are charged in a pay-as-you-go. Video stream providers utilize the cloud to stream videos to viewers. The major cloud services for video streaming are:
\begin{itemize}
\item Virtual Machines: are used for transcoding operations of videos and are charged per hour.
\item Storage: This is an essential service provided by cloud providers. It is a based monthly charge
\item Content Delivery Network (CDN): is a service in the cloud that offers fast access to contents like video streams, it avoids delay to stream videos. CDN replicates video streams in different geographical locations to reduce the traffic travel time in network ~\cite{saro,vakali}.
\end{itemize}
Amazon Web services (AWS)\footnote{https://aws.amazon.com/} is well known for cloud services and the great features with high reliability presented to users. Even though, our approach is not dependent on Amazon only and could be applied to any other cloud providers. In this research, we consider AWS for the evaluation of our proposed approach.

Different types of storage are provided by AWS:
\begin{itemize}
\item \texttt{S3 Standard}: It is typically used to store frequently accessed video streams.
\item \texttt{S3 Standard-Infrequent Access (S3 Standard-IA)}: is used to store infrequently accessed video streams.
\item \texttt{One Zone-Infrequent Access (S3 One Zone-IA)}: is used to store less infrequently accessed video streams.
\item \texttt{S3 Glacier}: stores frequently access video streams for the long term.
\end{itemize}

\section{Proposed Clustering Video Stream Method}\label{proposed}
Table \ref{symbols} represents the description of the symbols used in the equations and the proposed algorithm.
\begin{table}
  \centering 
  \begin{tabular}{cc}
\hline \hline
\multicolumn{1}{l}{ Symbols}  & \multicolumn{1}{l}{Description}   \\
 \hline

  \multicolumn{1}{l}{$m_i$} &\multicolumn{1}{l}{ Total number of GOPs in video $i$}\\
\multicolumn{1}{l}{  $G_{ij}$ }&\multicolumn{1}{l}{  $j^{th}$ GOP in video $i$} \\
\multicolumn{1}{l}{  $G_{threshold}$ }&\multicolumn{1}{l}{  GOP threshold in video $i$} \\
\multicolumn{1}{l}{  $j$ }&\multicolumn{1}{l}{  index  of the GOP $G_{ij}$ in video $i$} \\
  \multicolumn{1}{l}{$\gamma_i$ }& \multicolumn{1}{l}{ Views received by video $i$ }\\
 \multicolumn{1}{l}{ $\xi_{ij}$} & \multicolumn{1}{l}{Estimated views received by GOP $G_{ij}$}\\
  \multicolumn{1}{l}{ $S_{ij}$} & \multicolumn{1}{l}{Size of GOP $G_{ij}$ in megabytes}\\
 \multicolumn{1}{l}{ $P_{storage}$} &\multicolumn{1}{l}{  General price of cloud storage (\$/ GB)} \\
  \multicolumn{1}{l}{ $P_{storage_1}$} &\multicolumn{1}{l}{  Price of cloud storage \texttt{S3 Standard} used in cluster 1 (\$/ GB)} \\
   \multicolumn{1}{l}{ $P_{storage_2}$} &\multicolumn{1}{l}{  Price of cloud storage \texttt{S3 Standard-IA} used in cluster 2 (\$/ GB)} \\
 \multicolumn{1}{l}{ $P_{storage_3}$} &\multicolumn{1}{l}{  Price of cloud storage \texttt{S3 One Zone-IA} used in cluster 3 (\$/ GB)} \\
  \multicolumn{1}{l}{ $P_{storage_4}$} &\multicolumn{1}{l}{  Price of cloud storage \texttt{S3 Glacier} used in cluster 4 (\$/ GB)} \\
  \multicolumn{1}{l}{$P_{tran}$}&  \multicolumn{1}{l}{Price of cloud virtual machine for (\$/hour)} \\
  \multicolumn{1}{l}{$SC_{G_{ij}}$ }&\multicolumn{1}{l}{ Storage cost of GOP $G_{ij}$} \\
   \multicolumn{1}{l}{$TC_{G_{ij}}$} & \multicolumn{1}{l}{Transcoding cost of GOP $G_{ij}$}\\
  \multicolumn{1}{l}{$TC_{video_{i}}$} & \multicolumn{1}{l}{Transcoding cost of video $i$}\\
   \multicolumn{1}{l}{$SC_{video_{i}}$ }&\multicolumn{1}{l}{ Storage cost of video $i$} \\
     \multicolumn{1}{l}{$SC_1$ }&\multicolumn{1}{l}{ Storage cost of GOPs in cluster 1  } \\
      \multicolumn{1}{l}{$SC_2$ }&\multicolumn{1}{l}{ Storage cost of GOPs  in cluster 2  } \\
       \multicolumn{1}{l}{$SC_3$ }&\multicolumn{1}{l}{ Storage cost of GOPs in cluster 3  } \\
        \multicolumn{1}{l}{$SC_4$ }&\multicolumn{1}{l}{ Storage cost of GOPs in cluster 4  } \\
    \multicolumn{1}{l}{$ SC_{1-threshold}$ }&\multicolumn{1}{l}{ Storage cost of  first GOP till GOP threshold  } \\
       \multicolumn{1}{l}{$ TC_{remaining}$ }&\multicolumn{1}{l}{ Transcoding cost of  the remaining GOPs after the threshold } \\
   \multicolumn{1}{l}{$R_{G_{ij}}$} & \multicolumn{1}{l}{Ratio of storage cost to transcoding cost of GOP  of $G_{ij}$ }\\
   \multicolumn{1}{l}{$R_{video_i}$} & \multicolumn{1}{l}{Ratio of storage cost to transcoding cost of video  of $i$ }\\
    \multicolumn{1}{l}{$COST_{total}$} & \multicolumn{1}{l}{Total cost of using cloud for video  of $i$ }\\

\hline \hline
\end{tabular}
  \caption{Symbols used in the proposed method }
\label{symbols}
\end{table}

\subsection{Proposed Storage Approach}
\begin{figure}[H]
    \includegraphics[width=9cm, height=5cm]{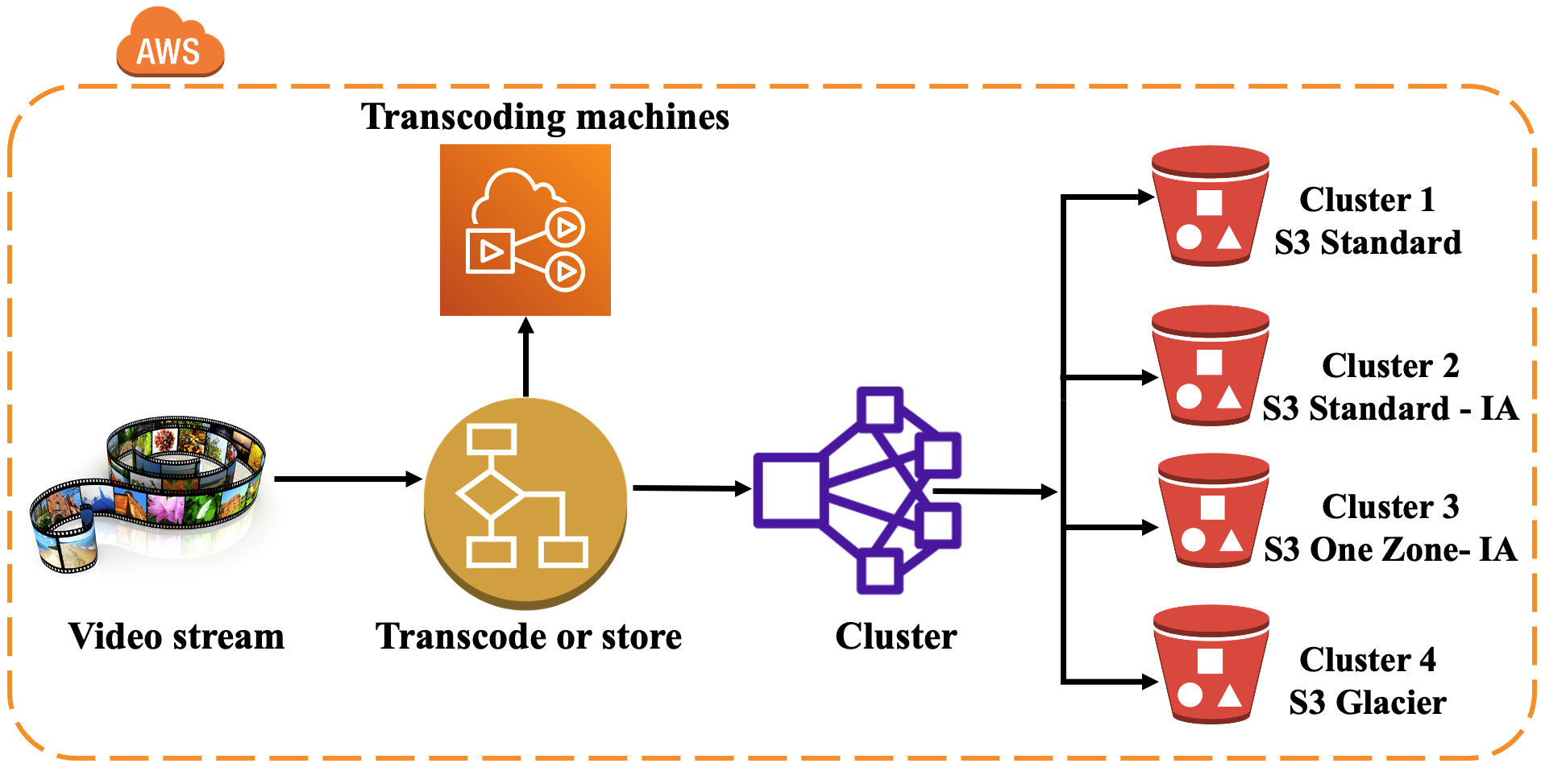}
    \caption{Proposed storage approach for video streams in cloud}
    \label {diagram}
\end{figure}

We propose an approach that facilitates the management of video storage in the cloud. Fig.\ref{diagram} presents the flow of video processing. In the cloud, a video-on-demand has two options either stored or transcoded (i.e., the video is deleted and transcoded upon request by the users). Once the video is selected to be stored, it goes through the clustering method to send it to the suitable storage. the storage mechanism is based on the number of views received by a video and the accessibility rate of the cloud storage. If a video doesn't meet the criteria for storage, it is deleted from the repository. In case a user requests it to watch, the original video, which is stored cloud, will be transcoded and generated in the right format for the video to fit the user's device.

\subsection{Video Streams Access Pattern}
 The cost of using cloud service to store or transcode video streams is mainly dependent on accessing video streams. Therefore, it is important to explore the pattern of video accesses. Previous studies revealed the access pattern to videos is a long-tail distribution ~\cite{miranda}. In such a pattern, the beginning of a video is highly watched and the rest parts are almost rarely accessed. Therefore, the distribution of GOPs in a video stream  could be implemented by applying the  Power Law model~\cite{newman} to represent the long-tail pattern.  Thus, the views of GOPs could be estimated.   Assume $G_{ij}$ is the $jth$ GOP in a video i and its views $\xi_{ij}$ is estimated in in Eq. \ref{gop-predict} where $\gamma_{i}$   represents the views of video $i$ and  $\alpha$ is constant and $j$ is the index of GOP in the video $i$. We select the value of $\alpha$ equals to $0.1$ to model the long-tail distribution of GOPs in the video .

\begin{equation}
\xi_{ij}=\dfrac{\gamma_{i}}{ j^{\alpha} }
\label{gop-predict}
\end{equation}

\subsection{ Cost of  Storing Video Streams in Cloud}

The rates of cloud storage are charged based on the size of a video. Note that a video stream is composed of many GOP. Knowing the size of a GOP $G_{ij}$, the storage cost of a $j^{th}$ in a video stream $i$ is computed in Eq.\ref{storage} where $SC_{G_{ij}}$ is the storage cost of $G_{ij}$, $S_{ij}$ is the size of $G_{ij}$ in megabytes, and  $P_{storage}$ is the cloud storage rate in  (\$/GB). In  Eq.\ref{storage}, we divide by $2^{10}$ to convert from megabytes to gigabytes. 

\begin{equation}
SC_{G_{ij}}=\dfrac{S_{ij} \cdot P_{storage}}{2^{10}}
  \label{storage}
\end{equation}
 Similarly, the storage cost of a whole video is calculated by summing up the storage cost of its all GOPs,  the concept of Eq.\ref{storage} is extended to a video level  and thus the storage cost of a video stream $i$ is $SC_{video_{i}}$ and  illustrated in   Eq.\ref{gopstorage} 
\begin{equation}
SC_{video_{i}}=\sum SC_{G_{ij}}=\sum\dfrac{S_{ij} \cdot P_{storage}}{2^{10}}
\label{gopstorage}
\end{equation}

\subsection{ Cost of Transcoding Video Streams in Cloud}\label{tc}
The cost of transcoding a video depends on the time of utilization of the virtual machine in the cloud and its type, the VM rate is in an hourly basis. Assume $TC_{G_{ij}}$ is the transcoding cost of GOP $G_{ij}$ in a video $i$, $\tau_{ij}$ is the time to complete the transcoding operation in seconds, and $P_{trans}$ is the price of a virtual machine per hour. The equation of the transcoding cost is given in Eq.\ref{trans}. We divide by 3600 to convert seconds to an hour.

\begin{equation}
TC_{G_{ij}}= \dfrac{ \tau_{ij} \cdot P_{trans}}{3600}
\label{trans}
\end{equation} 
It is worth to mention that Eq.\ref{trans} estimates the transcoding cost of GOP one time. Therefore, if the GOP $G_{ij}$  receives $\xi_{ij}$ views (i.e., which was estimated in Eq.\ref{gop-predict}), then  the transcoding cost of $G_{ij}$ is computed in Eq.\ref{costviews}
\begin{equation}
TC_{G_{ij}}= \dfrac{\xi_{ij} \cdot \tau_{ij} \cdot P_{trans}}{3600}
\label{costviews}
\end{equation}
It is noteworthy that the transcoding time $\tau_{ij}$ of GOP $ G_{ij}$ is based on the historical calculated transcoding time of the same GOP \footnote{For a video on demand VOD, we assume that  GOPs have been accessed by users, thus the transcoding time was computed before. In case of live streaming,  the transcoding time of GOPs could not be based on historical information.}

Accordingly, Eq.\ref{trans} is implemented at the video level to calculate the transcoding cost $TC_{video_{i}}$ for video stream $i$ by summing up the transcoding of its GOPs and is illustrated in Eq.\ref{videotrans}  assuming that GOP $G_{ij}$ in the video received $\xi_{ij}$ views which was estimated in Eq.\ref{gop-predict}
\begin{equation}
TC_{video_{i}}= \sum TC_{G_{ij}}=\sum \dfrac{\xi_{ij} \cdot \tau_{ij} \cdot P_{trans}}{3600}
\label{videotrans}
\end{equation}

\subsection{Decision of transcoding and storing a video stream}\label{decision}
Calculating  the cost of transcoding and storing of a GOP in the previous section, we can provide a decision and determine for each GOP in a video if it should be deleted and transcoded upon request or stored.
Therefore, the ratio of storage cost to transcoding cost of GOP $G_{ij}$ in a video stream $i$ is calculated and  is denoted as $R_{G_{ij}}$. The ratio is illustrated in Eq.\ref{ratio}.
\begin{equation}
R_{G_{ij}}=\dfrac{SC_{G_{ij}}}{TC_{G{ij}}}=\dfrac{(\dfrac{S_{ij} \cdot P_{storage}}{2^{10}})}{(\dfrac{\xi_{ij} \cdot \tau_{ij} \cdot P_{trans}}{3600})}
\label{ratio}
\end{equation}

When the storage cost $SC_{G_{ij}}$ of a GOP $G_{ij}$ is greater than its transcoding cost $TC_{G_{ij}}$ (i.e. $SC_{G_{ij}} >TC_{G_{ij}}$) it is worth to delete the GOP and transcode it upon request. In this case the ratio $R_{G_{ij}}>1$. That means the GOP $G_{ij}$ receives less views.

When the storage cost $SC_{G_{ij}}$ of a GOP $G_{ij}$ is less than its transcoding cost $TC_{G_{ij}}$ (i.e. $SC_{G_{ij}} <TC_{G_{ij}}$)  it is worth to store the GOP. In this case the ratio $R_{G_{ij}}<1$. That means the GOP $G_{ij}$ receives more views and becomes frequently accessed GOP.

When the storage cost $SC_{G_{ij}}$ of a GOP $G_{ij}$ is equal to its transcoding cost $TC_{G_{ij}}$ (i.e. $SC_{G_{ij}} =TC_{G_{ij}}$)  it is worth to store the GOP because it is easier to provide the user a stored GOP than transcode it on fly. In this case the ratio $R_{G_{ij}}=1$.

Similarly, the ratio concept discussed for GOPs could be achieved on video level. Thus, the ratio of storage cost to transcoding cost of video stream $i$ is illustrated in Eq.\ref{videoratio}.
\begin{equation}
R_{video_i}=\dfrac{SC_{video_{i}}}{TC_{video_{i}}}= \dfrac{\sum SC_{G_{ij}}}{\sum TC_{G_{ij}}}=\dfrac{(\sum\dfrac{S_{ij} \cdot P_{storage}}{2^{10}})}{(\sum \dfrac{\xi_{ij} \cdot \tau_{ij} \cdot P_{trans}}{3600})}
\label{videoratio}
\end{equation}

When the storage cost $SC_{video_{i}}$ of a video $i$ is greater than its transcoding cost $TC_{video_{i}}$ (i.e. $SC_{video_{i}} >TC_{video_{i}}$) it is worth to delete the video and transcode it upon request. In this case the ratio $R_{video_{i}}>1$. That means the video $i$ receives less views.

When the storage cost $SC_{video_{i}}$ of a video $i$ is less than its transcoding cost $TC_{video_{i}}$ (i.e. $SC_{video_{i}} <TC_{video_{i}}$)  it is worth to store the video. In this case the ratio $R_{video_{i}}<1$. That means the video $i$ receives more views and becomes frequently accessed video.

When the storage cost $SC_{video_{i}}$ of a video $i$ is equal to its transcoding cost $TC_{video_{i}}$ (i.e. $SC_{video_{i}} =TC_{video_{i}}$)  it is worth to store the video because it is easier to provide the user a stored version than transcode it on fly. In this case the ratio $R_{video_{i}}=1$.

\subsection{Algorithm \ref{al2}}\label{algorithm}

The algorithm is an extended version of the previous research~\cite{darwich20}. The purpose of the algorithm is to cluster the frequently accessed video stream and store them in different cloud storage to minimize the incurred cost and delete the non frequently accessed videos and transcode upon request using an original version. The algorithm is carried out on all GOPs in any video in the repository of video streams providers and its pseudo-code is shown in Algorithm \ref{al2}.

The inputs of the algorithm are GOPs, their sizes,  their transcoding times, price of cloud storage, and price of transcoding machines. The algorithm output decides if videos should be deleted and transcoded upon request  or stored in different cloud storage and then it computes the total cost of using cloud services for each video.

\begin{figure}[H]
 \includegraphics[width=9cm, height=5cm]{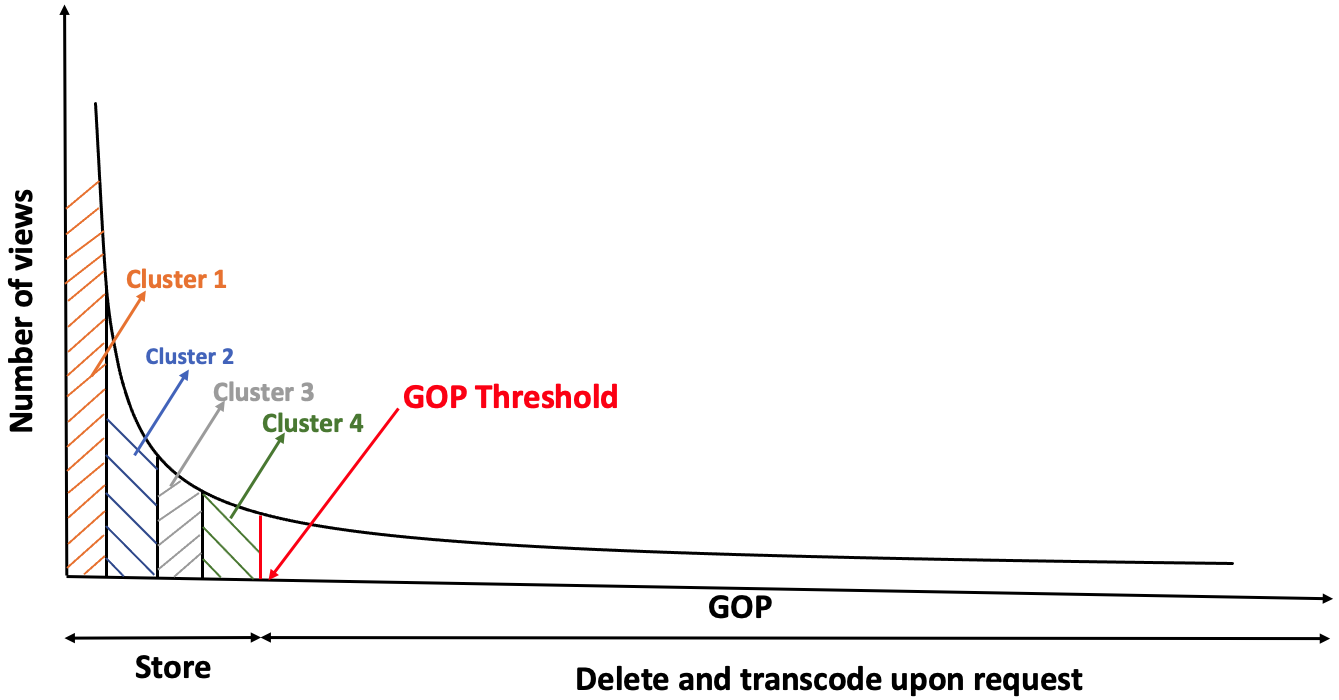}
    \caption{
    Access pattern to a video follows long-tail. All GOPs before $GOP_{threshold}$ are clustered and stored while the rest GOPs are deleted and transcoded upon request}
    \label {Video-partial4}
\end{figure}

For each GOP $G_{ij}$ in each video $i$ in the repository (Step 1), the algorithm estimates the views $\xi_{ij}$ received by each GOP by implementing Eq.\ref{gop-predict} (Step 2). The cost of storage and transcoding are computed by applying the equations ~\ref{storage} and \ref{trans} (steps 3 and 4). Then in the next step (Step 5), the ratio of storage cost to transcoding cost is calculated. If the ratio is less than 1, it means that the storage cost is less than the transcoding cost and hence  the algorithm decides to store the GOP (Steps 6 and 7). The algorithm keeps searching for a GOP that the next GOP has a ratio greater than 1. This GOP is called threshold GOP $G_{threshold}$.  All the remaining GOPs after the threshold have a ratio greater than 1 which means the storage cost is higher than the transcoding cost. The algorithm requests to delete all GOPs after the threshold and transcode them upon request (Steps 8 and 9). Fig.\ref{Video-partial4} illustrates the access pattern of GOPs in a video and which is a long-tail distribution. It is noticed that the beginning GOPs are more accessed than the rest GOPs which are located after the threshold GOP.
{\SetAlgoNoLine%

\begin{algorithm}
    \SetKwInOut{Input}{Input}
    \SetKwInOut{Output}{Output}

    \Input{\\  All $n$ $GOP$ of  video $i$\\
    Size of  GOP  $G_{ij}$ in video $i$: $S_{ij}$ \\
    
  Needed time to transcode GOP $G_{ij}$: $\tau_{ij}$\\
Cloud  Price of transcoding machine: $P_{trans}$\\
  Cloud Price of storage: $P_{storage}$\\
  Count views of GOP $G_{ij}$ : $\xi_{ij}$\\
 
  }
    \Output{Total cost of using cloud services for video stream $i$}
   For each GOP $G_{ij}$ in each video stream $i$\\
     
  \Indp    
   Estimated views for GOP $G_{ij}$: $\xi_{ij}\leftarrow \dfrac{\gamma_{i}}{ j^{\alpha} }$ \\

   Estimate cost of transcoding of GOP $G_{ij}$ that receives $\xi_{ij}$ views: $TC_{G_{ij}} \leftarrow \dfrac{\xi_{ij} \cdot \tau_{ij} \cdot P_{trans}}{3600}$ \\
   Compute cost of storage of GOP $G_{ij}$:  $SC_{G_{ij}}. \leftarrow \dfrac{S_{ij} \cdot P_{storage}}{2^{10}}$\\
    Compute the ratio: $R_{G_{ij}} \leftarrow \dfrac{SC_{G_{ij}}}{TC_{G{ij}}}$\\
       
     \eIf{$R_{G_{ij}}<=1$}
    {  Store $G_{threshold}\leftarrow G{_{i1}}$\\
    }
    { Delete and Transcode the remaining GOPs upon request;}
    \Indm
     
     Apply K-Means clustering on the stored GOPs   $ G_{i1}$ to $G_{threshold}$  with $K=4$\\
 Compute  storage cost of  Cluster 1: $SC_1 \leftarrow \sum  \dfrac{S_{ij} \cdot P_{storage_1}}{2^{10}}$\\
 Compute  storage cost of   Cluster 2: $SC_2 \leftarrow \sum  \dfrac{S_{ij} \cdot P_{storage_2}}{2^{10}}$\\
  Compute  storage cost of  Cluster 3: $SC_3 \leftarrow \sum  \dfrac{S_{ij} \cdot P_{storage_3}}{2^{10}}$\\
  Compute   storage cost of  Cluster 4: $SC_4 \leftarrow \sum  \dfrac{S_{ij} \cdot P_{storage_4}}{2^{10}}$\\
     Compute storage cost of   $ G_{1i}$ to $G_{threshold}$: $SC_{1-threshold}\leftarrow SC_1+ SC_2+ SC_3 + SC_4 $\\
     Compute   transcoding of the remaining GOPs in video $i$: $TC_{remaining} \leftarrow \sum \dfrac{\xi_{ij} \cdot \tau_{ij} \cdot P_{trans}}{3600}$\\
     Total cost of using cloud services for a video stream $i$:  $COST_{total} \leftarrow SC_{1-threshold}+TC_{remaining}$

    \caption{Clustering Method}
    \label{al2}
    
\end{algorithm}
}

Then the K-means algorithm is implemented with $K=4$ (steps 11-15) to cluster the stored GOPs. We select the value of $k=4$ to match the storage types in the cloud. The reason we rely on K- means is because of its simplicity and we do not consider many factors in the clustering process. Cluster 1 contains the most frequently accessed GOPs. Sequentially, clusters 2, 3, and 4 are dedicated for less frequently accessed GOPs. Note that the storage price of cluster 1 is higher than the rest of the cloud storage because it provides the fastest accessibility. While clusters 2, 3, and 4 have different prices and accessibility rates. In this research, we dedicate \texttt{S3 Standard } for Cluster 1, \texttt{S3 Standard-IA } for Cluster 2, \texttt{S3 One Zone-IA } for Cluster3, and \texttt{S3 Glacier} for cluster 4.  
In Step 16, the algorithm computes the storage cost of the stored GOPs. While in Step 17, it computes the transcoding cost of the remaining GOPs. The last Step 18, the total cost of using the cloud service, which is the summation of storage and transcoding costs,  is calculated  for the video $i$.

\section{Experiment Setup }\label{experiment}

\subsection{Videos Synthesis}

 Large repositories of video streams are necessarily required for the experiments. However, providing access to video stream providers is almost impossible. Furthermore, downloading video from the internet is a lengthy and costly procedure. Therefore, we come up with a way to synthesize video streams. The content of the synthesized videos is the characteristics that fit the experimental purposes. Particularly, size, transcoding time, and count of GOPs in a video are the necessary variables for the experiments.
 
\begin{figure}[H]
    \centering
        \includegraphics[width=0.55\textwidth]{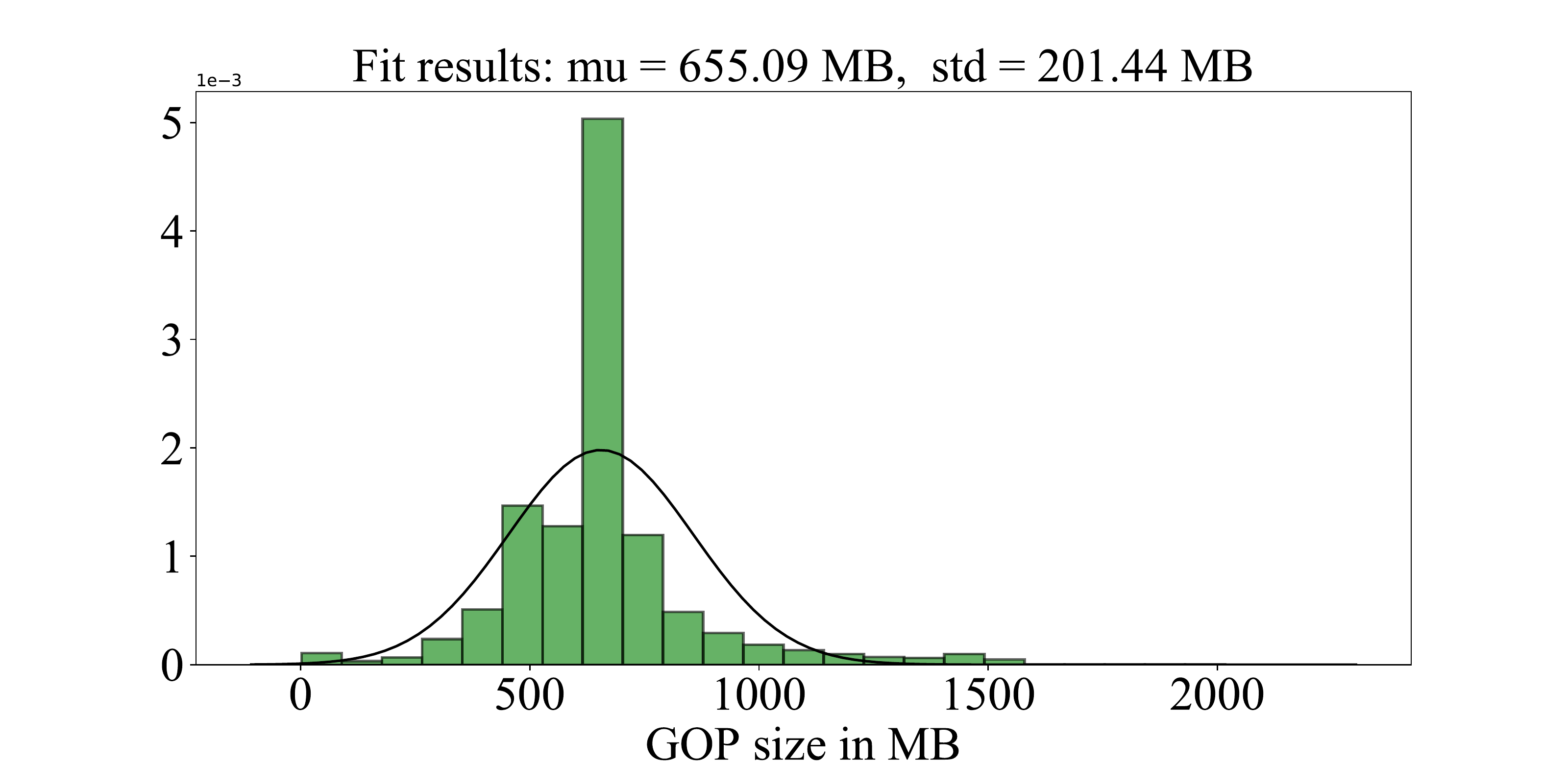}
        \caption{ Histogram of GOP size in a video }
        \label{fig:gopsizepdf}
    \end{figure}

\begin{figure}[H]
   \centering
        \includegraphics[width=0.55\textwidth]{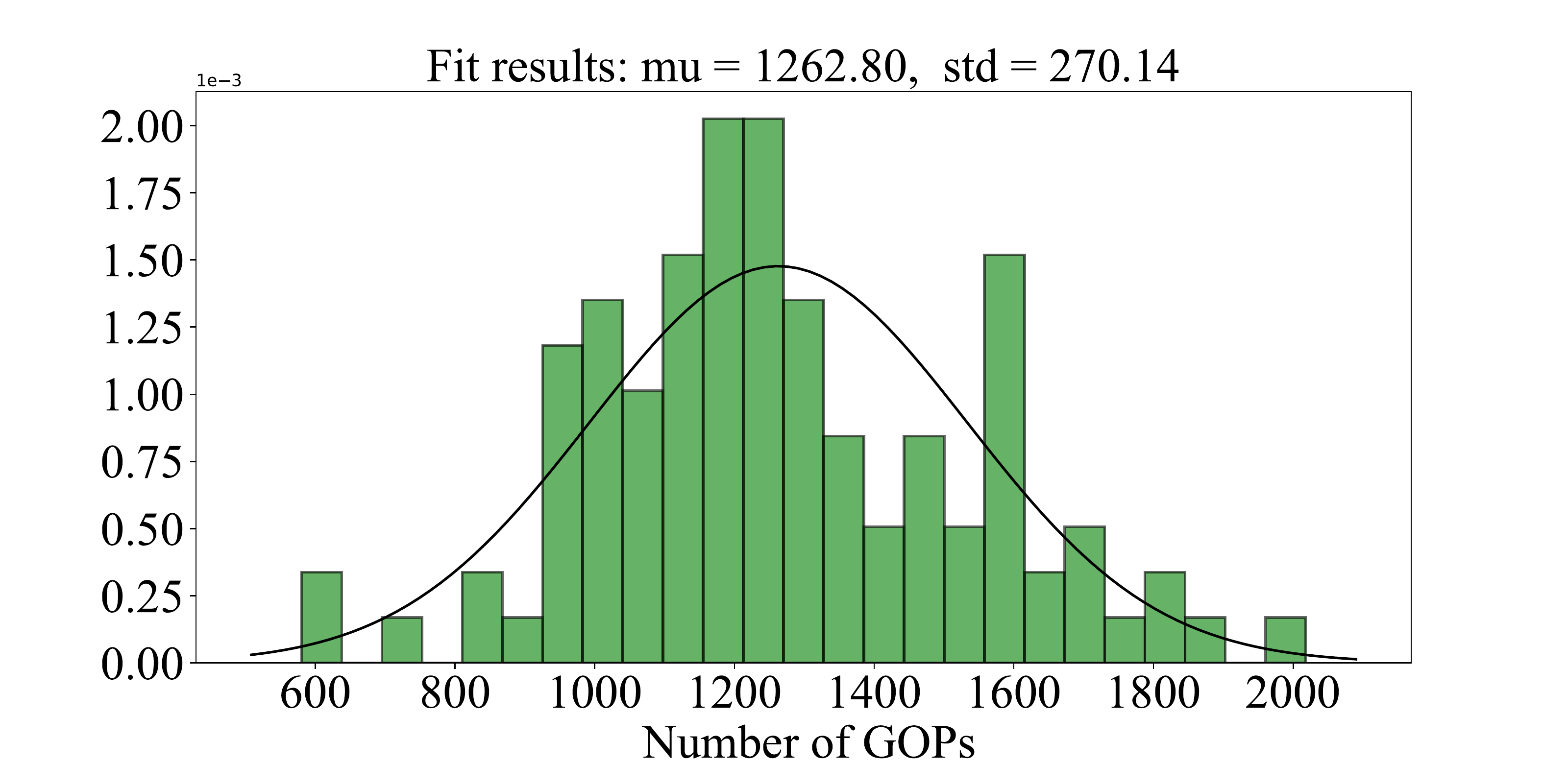}
        \caption{Histogram of GOPs count in a video}
        \label{fig:goppdf}
        \end{figure}

Accordingly, an analysis of a bunch of videos downloaded from YouTube is performed to reveal the relations among size, transcoding time, and numbers of GOPs in videos. The downloaded videos contain different categories including music, movies, animation, documentary, news, and sports. Transcoding operations are applied to the videos to generate information about the size, transcoding time, and GOPs number in each video. Furthermore, The extracted information is analyzed, we conclude that the GOPs sizes and GOPs number fits in a Gaussian curve as shown in Fig.~\ref{fig:gopsizepdf} and \ref{fig:goppdf}. The Gaussian values are detailed in table~\ref{statistics}. More importantly, we implement the linear regression analysis on GOPs size and transcoding time used in \cite{tech}, the analysis results in a linear equation between GOPs size and their corresponding transcoding time. The regression analysis is displayed Fig.~\ref{fig:regres}. 

By utilizing the generated data and the relationship between the variables, we create repositories of videos, each one contains 50,000 videos.

 \begin{table}[H]
\centering
\begin{tabular}{c c c c c}
\hline \hline
 &GOP size (Kbytes) & number of GOPs\\
\hline
  Mean   &655.08 & 1262.79   \\
   Standard Deviation & 201.44  & 271.46    \\
   Standard error mean & 0.57 & 26.74 \\
   Upper 95\% Mean & 656.20 & 1315.85\\
   Lower 95\% Mean & 653.96 & 1209.74\\
   Count of GOPs & 124593 & 130068 \\
   Min. size, count GOP &1.91 & 580 \\
   Max. size, count GOP & 2192.65 & 2018\\
   \hline \hline
     
   \end{tabular}
   \caption{ Data statistics  of GOPs size  and  their count  in a video}
   \label{statistics}
   \end{table}

\begin{figure}[H]

       
    \caption{Finding a relationship between GOP size and transcoding time by applying Linear regression analysis}
    \label{fig:regres}
\end{figure}

\subsection{ Videos Access Pattern} \label{view-model}
We create the access pattern to synthesized video in the repository by implementing a Weibull distribution equation. Accordingly, the plotted curve follows a long tail shape. The access pattern shows the frequently accessed videos at the beginning of the curve while the videos having fewer views are located in the curve's long tail. The Weibull distribution has two coefficients shape $\alpha$ and scale $\beta$ which play a critical role to define the percentage of frequently accessed videos in the repository. Decreasing the coefficient $\alpha$ value results in increasing the number of frequently accessed videos, we vary $\alpha$ between 0.4 and 2.4 and set $\beta=1$ in the experiments for the sake of having a long tail shape.

\subsection{Storage and Computation Prices}
The prices of  cloud storage and transcoding virtual machines utilized in the experiments are the actual prices of Amazon Web Service (AWS)  as detailed in Table~\ref{tab:price}
\begin{table}[H]
\centering 
\begin{tabular}{c c c c c} 
\hline\hline 
  Price Transcoding Machine  & Price CDN\\ [0.5ex] 
\hline 
\texttt{t2-small} & \texttt{CloudFront}\\
\$0.026  /hour & \$0.085  GB/month \\ 
\hline \hline 
\end{tabular}

\caption{Actual prices of AWS transcoding machines and content delivery network (CDN) in US dollar \$}
\label{tab:price}
\end{table}

 \begin{table}[H]
\centering 
\begin{tabular}{c c c c} 
\hline\hline 
  Storage Type  &  Storage Price \\ [0.5ex] 
\hline 

\texttt{S3 Standard } & \$0.023   GB/month  \\
\texttt{S3 Standard-IA } & \$0.0125   GB/month \\
\texttt{S3 One Zone-IA } &  \$0.01   GB/month\\
\texttt{S3 Glacier } & \$0.001   GB/month\\

\hline
\hline 
\end{tabular}

\caption{AWS  storage services and  prices in US dollar \$}
\label{tab:storageprice}
\end{table}

\subsection{ Proposed Method and other Methods for Comparison}
The proposed approach is compared to previous works and two baselines methods:
\begin{itemize}
 \item \emph{Fully Storing Videos Method} which stores all videos regardless of their frequency of access. 
 
\item  \emph{Deleting  and Transcoding all Videos Method}  them upon request.
 
\item   \emph{Patrtial-Storing videos} method \cite{darwich2016}  which stores the most-watched part of a video in one of the cloud storage.
  
\item   \emph{VideoTranscode or Delete Method} : This method reduces the cost of cloud services for video streams and was proposed by Jokhio \etal \cite{jokhio}. They proposed an approach to store the whole video or delete and transcode it upon request. They considered the popularity of videos in their method to decide which video should be stored or transcoded.
  
\item   \emph{Trade-off Between Storing and Transcoding Method} : Zhao \etal proposed a method to tare-off between storing and transcoding costs. Thus, their scheme reduced the cost of cloud services. In their approach, they used the transcoding relationships among the versions of a video with the views of each video. Their method decides which video versions should be stored or deleted and transcoded on demand.
\end{itemize}
\section { Experimental Results}\label{results}
\subsection{Effect of varying the percentage of frequently accessed videos in repository}
We build several repositories made of synthesized videos. Each repository contains 50,000 videos with a different percentage of frequently accessed videos. We adjust the shape coefficient in the Weibull distribution equation to change the percentage of the frequently accessed video in each repository. Table~\ref{FAVs} displays the $\alpha$ values with the corresponding percentage of frequently accessed videos.

\begin{center}
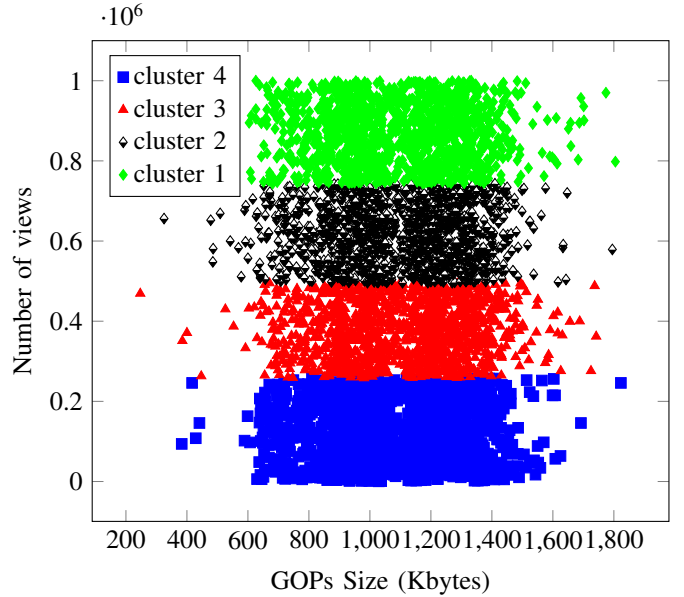


\captionof{figure}{Clustered frequently accessed GOPs based on the number of views}
\label{gop_clusters}
\end{center}

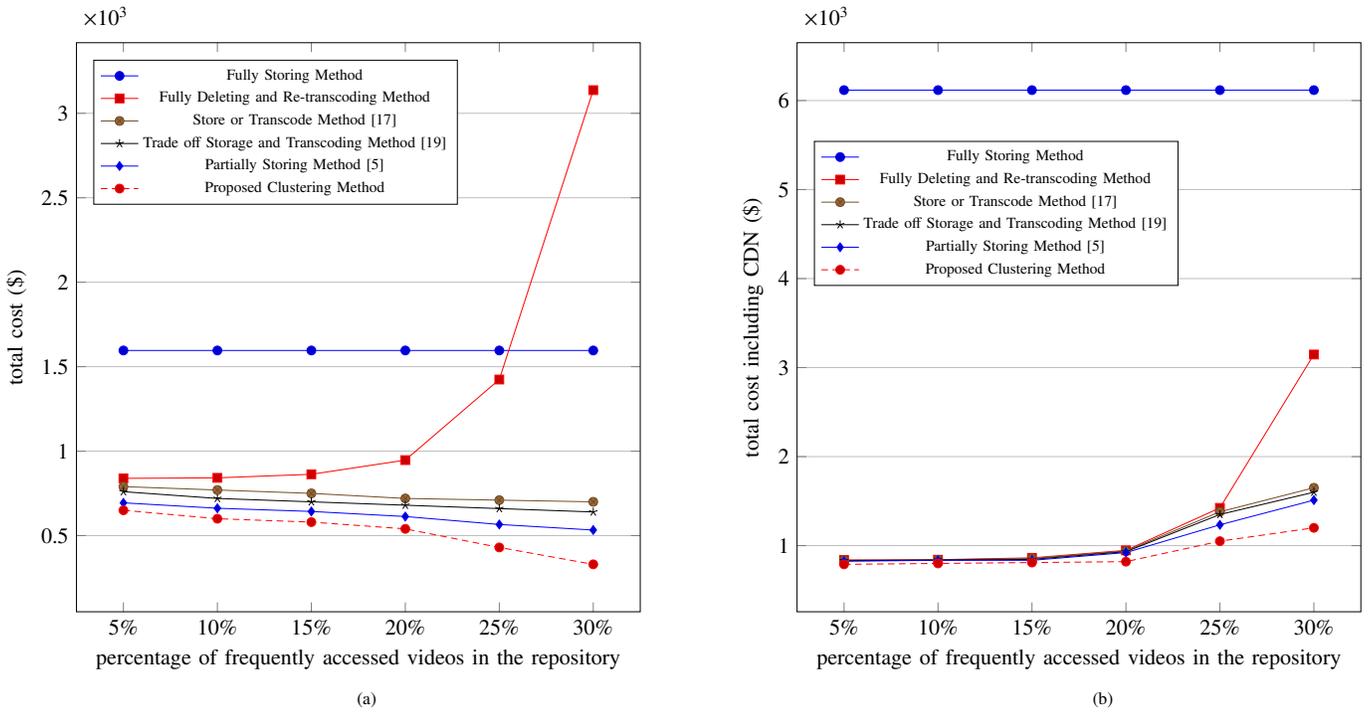
\begin{figure*}
\resizebox{2.21\columnwidth}{!}{%
 \begin{subfigure}[b]{0.65\textwidth}
\begin{tikzpicture} 
    \begin{axis}[
         width  =0.9*\textwidth,
        height = 10.7cm,
        ymajorgrids = true,
       legend style={font=\fontsize{7}{5}\selectfont},
        ylabel = {total cost (\$)},
        xlabel={percentage of frequently accessed videos in the repository},
          xticklabel={$\pgfmathprintnumber{\tick}\%$},
        xtick = data,
       scaled y ticks=real:1000,
    	ytick scale label code/.code={$\times10^3$},
        legend pos= north west,
           ]
       \addplot
            coordinates {(5, 1596)(10, 1596)(15, 1596)(20, 1596)(25, 1596)(30, 1596)};

  \addplot
               coordinates {(5, 839)(10, 842)(15, 863)(20, 947)(25, 1424)(30, 3137)};
                \addplot
               coordinates {(5, 790)(10, 770)(15, 750)(20,720)(25, 710)(30, 700)};
                \addplot
               coordinates {(5, 760)(10, 720)(15, 700)(20,680)(25, 660)(30, 640)};

      \addplot
              coordinates {(5, 694)(10, 662)(15, 643)(20, 613)(25, 566)(30, 533)};

 \addplot
              coordinates {(5, 650)(10, 600)(15, 580)(20, 540)(25, 430)(30, 330)};

        \legend{Fully Storing Method, Fully Deleting and Re-transcoding Method, Store or Transcode Method \cite{jokhio},Trade off Storage and Transcoding Method \cite{zhao}, Partially Storing Method \cite{darwich20}, Proposed Clustering  Method }
         
    \end{axis}
   
\end{tikzpicture}

\caption{}
\label {favs}
\end{subfigure}%

 \begin{subfigure}[b]{0.65\textwidth}

\begin{tikzpicture}
    \begin{axis}[
         width  = 0.9*\textwidth,
        height = 10.7cm,
        ymajorgrids = true,
       legend style={font=\fontsize{7}{5}\selectfont},
        ylabel = {total cost including CDN (\$)},
          xlabel={percentage of frequently accessed videos in the repository},
        xticklabel={$\pgfmathprintnumber{\tick}\%$},
        xtick = data,
         scaled y ticks=real:1000,
    	ytick scale label code/.code={$\times10^3$},
       legend style={at={(0.03,0.7)},anchor=west},
           ]
       \addplot
            coordinates {(5, 6117)(10,  6117)(15,  6117)(20,  6117)(25,  6117)(30,  6117)};

  \addplot 
             coordinates {(5, 839)(10, 842)(15, 863)(20, 947)(25, 1424)(30, 3147)};
\addplot 
             coordinates {(5, 835)(10, 840)(15, 855)(20, 940)(25, 1380)(30, 1650)};

\addplot 
             coordinates {(5, 830)(10, 838)(15, 845)(20, 932)(25, 1350)(30, 1600)};
      \addplot
                  coordinates {(5, 825)(10, 835)(15, 835)(20, 922)(25, 1234)(30, 1511)};
       
  \addplot
            coordinates {(5, 790)(10, 800)(15, 810)(20,820)(25, 1050)(30, 1200)};
        \legend{Fully Storing Method, Fully Deleting and Re-transcoding Method, Store or Transcode Method \cite{jokhio},Trade off Storage and Transcoding Method \cite{zhao}, Partially Storing Method \cite{darwich20},  Proposed Clustering  Method}
         
    \end{axis}
   
\end{tikzpicture}

\caption{}
\label {cdn}
\end{subfigure}%

}
\caption{Incurred costs of Fully Storing, Fully Deleting and Re-transcoding,  Store or Transcode \cite{jokhio},Trade off Storage and Transcoding \cite{zhao}, Partially Storing \cite{darwich20}, and Proposed Clustering   methods  methods: (\ref{favs})  percentage of frequently accessed videos is changed in the repository; (\ref{cdn})  Content delivery network cost is considered to the storage cost;  }
\end{figure*}

\begin{table}[H]
\centering 
\begin{tabular}{|c| c c|} 
\hline
\multirow{6}{*} {$\beta=1$}&$\alpha$ &     Frequently Accessed Videos  \\ [0.5ex]
\hline
& 0.4 &30\% \\ 
& 0.6   & 25\% \\ 
 & 1 & 20\% \\
 & 1.4 & 15\% \\
 & 1.8 & 10\% \\
 & 2.4 & 5\% \\
\hline

\end{tabular}
\label{price} 
\caption{Percentage of frequently accessed videos is changed by adjusting $\alpha$ in the Weibull distribution equation}
\label{FAVs}
\end{table}
The simulation results are presented in Fig.\ref{favs}. The horizontal axis represents the varying percentage of frequently accessed videos and the vertical axis represents the total cost of cloud services. We note that the cost of applying the fully storing method is around \$1600 and constant along with a different percentage of frequently accessed videos because it aims to store all videos regardless of the frequency access to the videos. The method of deleting videos and transcoding upon request work fairly for a small percentage of frequently accessed videos. However, it incurs a high cost when the percentage gets increased because we assume each time the user views the video, we apply the transcoding operation on the original video.

\begin{figure}
\resizebox{0.9\columnwidth}{!}{%
\begin{tikzpicture}
    \begin{axis}[
         width  = 0.51*\textwidth,
        height = 10.7cm,
        major x tick style = transparent,
        ymajorgrids = true,
       legend style={font=\fontsize{6}{5}\selectfont},
        ylabel = {total cost (\$)},
        xlabel={ Increased percentage of views count in a video },
        symbolic x coords={1,2,3,4,5 },
          xticklabel={$\pgfmathprintnumber{\tick}\%$},
        xtick = data,
         scaled y ticks=real:1000,
   		 ytick scale label code/.code={$\times10^3$},
        legend pos= north west,
           ]
        \addplot
            coordinates {(1, 1596)(2, 1596)(3, 1596)(4, 1596)(5, 1596)};

  \addplot
               coordinates {(1, 947)(2, 1112)(3, 1451)(4, 1834)(5, 2234)};
               
                \addplot
              coordinates {(1, 850)(2, 650)(3,550)(4, 510)(5, 470)};
                \addplot
              coordinates {(1, 800)(2, 600)(3,500)(4, 470)(5, 440)};

      \addplot
              coordinates {(1, 769)(2, 543)(3,470)(4, 436)(5, 418)};

 \addplot
              coordinates {(1, 730)(2, 500)(3,410)(4, 370)(5, 330)};

        \legend{Fully Storing Method, Fully Deleting and Re-transcoding  Method,  Store or Transcode Method \cite{jokhio},Trade off Storage and Transcoding Method \cite{zhao}, Partially Storing Method \cite{darwich20}, Proposed Clustering  Method }

    ]
         
    \end{axis}
   
\end{tikzpicture}%
}
\caption{Incurred costs of Fully Storing, Fully Deleting and Re-transcoding,  Store or Transcode \cite{jokhio},Trade off Storage and Transcoding \cite{zhao}, Partially Storing \cite{darwich20}, and Proposed Clustering   methods   (\ref{views}) when the count views of videos  is raised. }
\label {views}%

\end{figure}
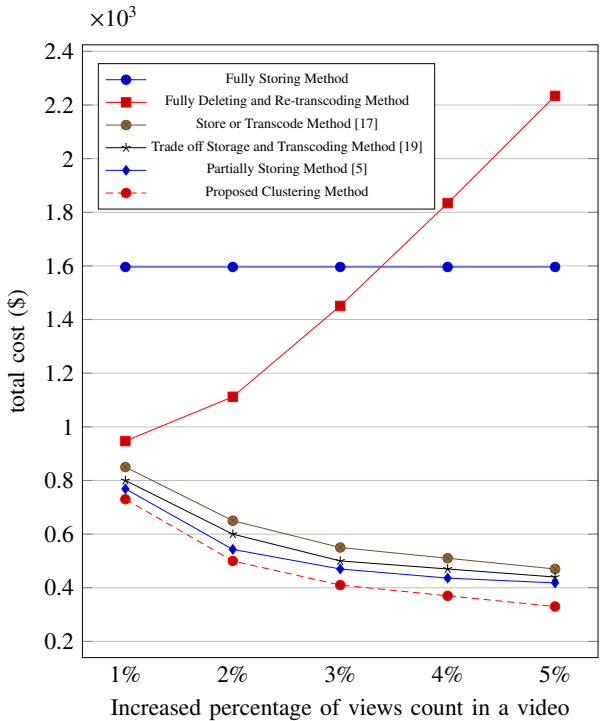

The store or transcode method \cite{jokhio} significantly minimizes the cost of using cloud service, it decides efficiently which videos streams should kept in the repository and which videos should be deleted and transcoded upon demand. Moreover, it could reduce the cost by up  57\% compared to fully storing method. We notice that the method steadily keeps the cost almost unchangeable and low when the percentage of the frequently accessed increases in the repository.

The trade-off storage and transcoding method \cite{zhao} further decreases the cost of cloud services. It highly reduces the cost up by 60\% compared to the fully storing method. This method has slightly more impact on reducing the cost when the percentage the frequently accessed video streams gets high in the repository.

It is worth to mention that methods of  \cite{jokhio} and  \cite{zhao}  are  essentially based on transcoding or storing the whole video. While the partially storing method \cite{darwich20}, which partially stores or transcode the GOPs of a video, outperforms the two aforementioned methods, it reduces the cost significantly as it stores only the frequently accessed parts of the videos in the S3 standard storage and deletes the rest. It could decrease the incurred cost by up to 75\% compared to the fully storing method. It is worth mentioning that the method works efficiently when the percentage of frequently accessed increases because it decides more parts of videos to be stored, which is less expensive than transcoding them. The proposed method outperforms all methods because it stores frequently accessed GOPs of each video in different cloud storage. It differs from the partially storing method, which stores all frequently accessed video in one storage (i.e. S3 standard), and which has the highest price, while the proposed method aims to distribute the videos in different cloud storage which have lower rates as shown in Fig.\ref{gop_clusters}. It could reduce the cost by 20\% compared to the partially storing method.

Similarly, we run our experiments using the Content Delivery Network shown in Fig.\ref{cdn} which incurs a high significant cost for fully storing method around \$6000 because all videos are replicated and stored in different geographical locations. As the percentage of frequently accessed videos increases, the cost of using CDN increases, since more replicated videos are stored in different locations. Our proposed method outperforms the other methods and reduces the cost by up to 15\% compared to the partially storing method.

\subsection{Effect of varying the number of views for frequently accessed videos}

The frequently accessed videos have variant views in repositories. It is worthwhile to study the effect of view variation when applying the proposed method. In this experiment, we increase the average views of frequently accessed videos by up to 5\%. The fully storing method keeps the cloud cost constant along with changing views because it stores all videos in the repository regardless of frequency of access and their views variations, this method imposes significant cost paid by the videos providers. The fully deleting and transcoding method increases the cloud cost gradually when the views rate increase because for each single view request, the video is transcoded and the cloud charges video stream providers. When the views of a video are increased by 5\% the cost is doubled as shown in Fig.\ref{views}. 

The two methods   \cite{jokhio} and  \cite{zhao} efficiently reduce the cost of cloud services. particularly, both methods significantly decreases the cost by up  69\% compared to the fully storing method. As the views of video streams increase, both methods work efficiently.

The partially storing method reduces the incurred cost significantly when videos views increase because more parts (i.e., GOPs) are stored in \texttt{S3 standard} storage which incurs less cost than transcoding. Moreover, it overcomes the performance of the aforementioned methods and decreases the cost further by up 7\% compared to the methods  \cite{jokhio} and  \cite{zhao}.

The proposed clustering method reduces further the cost and outperforms the other methods because its concept is based on storing the frequently accessed video in \texttt{S3 Standard},\texttt{S3 Standard-IA}, \texttt{One Zone-IA}, and \texttt{S3 Glacier} which charge less than \texttt{S3 Standard}. The overall cost reduction by implementing the proposed method is 20\% compared to the partially storing method.

\section{Conclusion}\label{conclusion}

In this paper, we propose a method to minimize the cost of cloud resources. This research is beneficial for video stream providers, it aims to minimize the cost of cloud resources. Particularly, we develop a method that clusters similar frequently accessed videos based on the views and then distributes them to suitable cloud storage. Our proposed method shows efficiency in cost reduction when the repository contains a high percentage of frequently accessed video, it reduces the cost by up to 20\% compared to the partially storing method. Moreover, when increasing the views of frequently accessed videos the incurred cost is reduced by up to 15\%. In addition, when considering CDN in the experiments, our proposed method outperforms other methods and decreases the cloud cost by up to 22\%.

\textbf{Funding}: No funding resources
\section{Compliance with Ethical Standards}

\textbf{Conflict of Interest} The authors declare that they have no conflict of interest.

%
\IEEEpeerreviewmaketitle

\ifCLASSOPTIONcaptionsoff
  \newpage
\fi



%

%

\begin{IEEEbiography}[{\includegraphics[width=1in,height=1.25in,clip,keepaspectratio]{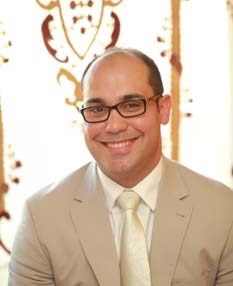}}]{Dr. Mahmoud Darwich}
 earned his bachelor’s degree in Electrical, Electronics and Communications Engineering from Jami’at Bayrut Al-Arabiya and his master’s and doctorate degrees in Computer Engineering from the University of Louisiana at Lafayette. He is currently an assistant professor of computer science in the Mathematical and digital sciences department at the Bloomsburg University of Pennsylvania. His research interests include cloud computing, video streaming, and VLSI design.
\end{IEEEbiography}

\begin{IEEEbiography}[{\includegraphics[width=1in,height=1.25in,clip,keepaspectratio]{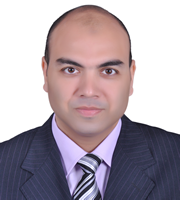}}]{ Dr. Yasser Ismail}
Dr. Yasser Ismail received his B.Sc.
degree in Electronics \& Communications
Engineering from Mansoura University -
Egypt, in 1999. He received his M.Sc. in
Electrical Communications from
Mansoura University - Egypt, in 2002.
Dr. Ismail received his M.S. and Ph.D.
degrees in Computer Engineering from
University of Louisiana at Lafayette -
USA in 2007 and 2010. Dr.
Ismail is currently working as an assistant professor in the Electrical
Engineering Department, Southern University and A\&M College -
Baton Rouge - Louisiana - USA. His area of expertise is Digital
Video Processing Algorithms/Architectures levels, Internet of
Things (IoT), VLSI and FPGA Design (Low-Power and High-Speed
Performance Embedded Systems), automotive transportation,
Robotics, RFID, and Wireless and Digital Communication Systems.
\end{IEEEbiography}


\begin{IEEEbiography}[{\includegraphics[width=1in,height=1.25in,clip,keepaspectratio]{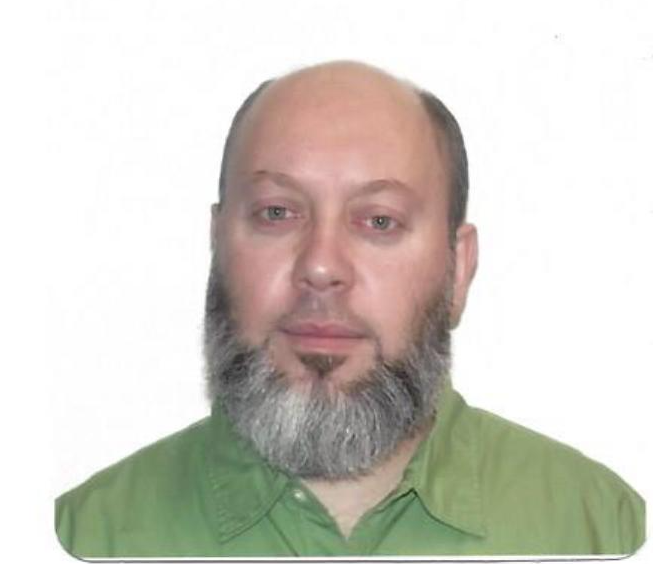}}]{ Dr. Talal Darwich}
 received the B.S. degree
in electrical engineering, communications
and electronics, from Beirut Arab University,
Lebanon in 1998, the M.S. degree and the Ph.D.
degree, in computer engineering from University
of Louisiana at Lafayette in 2002 and 2007 respectively. Dr. Darwich has wide experiences developing in-house CAD tools. He worked for Oracle, Intel, Xilinx, And currently is a staff lead at
Microchip Incorporation. His research covers developing in-house CAD tools for FPGA,
VLSI circuits design, low power circuits, timing, and signal integrity.
\end{IEEEbiography}
\begin{IEEEbiography}[{\includegraphics[width=1in,height=1.25in,clip,keepaspectratio]{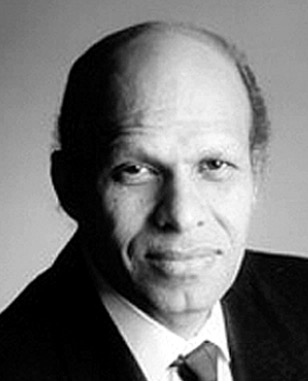}}]{Dr. Magdy Bayoumi}
 received B.Sc. and M.Sc. degrees in Electrical Engineering from Cairo University, Egypt; M.Sc. degree in Computer Engineering from Washington University, St. Louis; and Ph.D. degree in Electrical Engineering from the University of Windsor, Canada. He is presently the Z.L. Loflin Eminent Scholar Endowed Chair Professor at The Center for Advanced Computer Studies (CACS) and Director of Electrical Engineering Department, University of Louisiana at Lafayette (UL Lafayette). He was
the Vice President for Conferences of the IEEE
Circuits and Systems (CAS) Society. He is the
recipient of the 2009 IEEE Circuits and Systems
Meritorious Service Award and the IEEE Circuits
and Systems Society 2003 Education Award.
  \end{IEEEbiography}

\end{document}